\documentclass[11pt]{article}

\usepackage{jheppub}
\usepackage{amssymb}
\usepackage{amsmath}
\usepackage[title]{appendix}
\usepackage{graphicx}
\usepackage{graphics}
\usepackage{color}
\usepackage{fancyhdr}
\usepackage{graphicx}
\usepackage{float}
\usepackage{enumerate}
\usepackage{soul}

\usepackage[colorlinks=true,linktocpage=true,linkcolor=blue,citecolor=blue]{hyperref}

\usepackage[utf8]{inputenc}

\definecolor{darkgreen}{RGB}{50,150,0}

\title{Dark Dimension Gravitons as Dark Matter}

\author[a]{Eduardo Gonzalo,}
\author[c]{Miguel Montero,}
\author[c]{Georges Obied}
\author[c]{and Cumrun Vafa}

\affiliation[a]{Department of Physics, Lehigh University, Bethlehem, PA 18018, USA}
\affiliation[c]{ Department of Physics, Harvard University, Cambridge, MA 02138, USA}

\begin{document}

\abstract{
We consider cosmological aspects of the Dark Dimension (a mesoscopic dimension of micron scale), which has recently been proposed as the unique corner of the quantum gravity landscape consistent with both the Swampland criteria and observations.
In particular we show how this leads, by the universal coupling of the Standard Model sector to bulk gravitons, to massive spin 2 KK excitations of the graviton in the dark dimension (the ``dark gravitons'') as an unavoidable dark matter candidate.  Assuming a lifetime for the current de Sitter phase of our universe of order Hubble, which follows from both the dS Swampland Conjecture and TCC, we show that generic features of the dark dimension cosmology can naturally lead to the correct dark matter density and a resolution of the cosmological coincidence problem, where the matter/radiation equality temperature ($T\sim$ 1 eV) coincides with the temperature where the dark energy begins to dominate. 
Thus one does not need to appeal to Weinberg's anthropic argument to explain this coincidence.
The dark gravitons  are produced at $T\sim$ 4 GeV, and their composition changes as they mainly decay to lighter gravitons, without losing much total mass density.  The mass of dark gravitons is $m_{\text{DM}}\sim 1-100$ keV today.
}

\maketitle

\section{Introduction}

 In a recent paper~\cite{Montero:2022prj} it was shown that Swampland criteria combined with experimental observations, lead to a specific corner of the quantum gravity landscape.  In particular the smallness of the cosmological constant $\Lambda$, combined with the generalized Distance Conjecture~\cite{Ooguri:2006in,Lust:2019zwm}, the Emergent String Conjecture \cite{Lee:2019wij} and observational data lead to the prediction of exactly one mesoscopic dimension of radius in the micron range whose length is set by the dark energy length scale $\Lambda^{-1/4}$.  In this scenario the matter fields arise from branes localized in the mesoscopic dimension.  

In this paper we study the cosmological aspects of this model.  In particular we show that a natural and unavoidable component of the dark sector is dark graviton: the spin 2 massive KK excitations of the graviton in the dark dimension.  The basic setup is more or less forced on us by the fact that we have one mesoscopic dimension and the observed lack of light degrees of freedom in equilibrium with visible sector, as in the LED scenarios \cite{Arkani-Hamed:1998sfv}. Taken together, these lead us to assume that the Standard Model brane is in a thermal state at some initial temperature $T_i$, while the bulk modes living in the mesoscopic dimension remain essentially unexcited. As the Standard Model brane begins cooling off, its universal coupling to dark gravitons leads to  unavoidable KK graviton production\footnote{This fact was already noted in \cite{Arkani-Hamed:1998sfv}, and precisely for this reason it was pointed out that KK gravitons would have been an ideal candidate for the dark matter.  But as they note the range of parameters that are forced on LED scenarios to bring down the Planck mass to weak scale does not allow an accounting of dark matter using KK graviton. This is unlike the dark dimension scenario which is motivated instead by the dark energy hierarchy, and where the KK modes are a viable dark matter candidate.}. After production, the massive dark graviton
excitations decay predominantly to lighter gravitons (as there are more channels available compared to SM brane modes)
and quickly lower their mass distribution, without losing much total mass due to relative smoothness of the dark dimension (we assume structure vary not much smaller than $\sim 10^{-2} l$ where $l$ is the length of the dark dimension).  Using their decay we can estimate the evolution of dark matter over the age of the universe leading surprisingly to no conflict with any experimental observations.  The dark gravitons get produced at a mass of about $\sim 1-50$ GeV and the bulk of their mass shifts down to about 1-100 keV today.  This leads to a particular realization of the dynamical dark matter scenario proposed in \cite{Dienes:2011ja}. 

It is already interesting that for some value of the initial temperature, this model is compatible with observations. But we can do better. Using the dS Swampland conjecture combined with the requirement that the internal dimensions should be sufficiently stabilized compared to the dS scales (otherwise, the cosmology would not be four-dimensional), leads us to the natural scale for the initial temperature: 
\begin{equation}T_i \sim M_p \left({\frac{\Lambda}{ M_p^d}}\right)^{\frac{1}{2(d-1)}}\sim 1\, \text{GeV\quad  for $d=4, \Lambda\sim 10^{-120} M_p^4$}.\end{equation}   

The dark gravitons of this mass scale will initially constitute the bulk of the dark matter in our universe and begin decaying predominantly to lighter dark gravitons, without losing much total mass.  Remarkably this automatically gives the correct density of the observed dark matter in our universe!  In particular this scenario leads to the resolution of the cosmological coincidence problem\footnote{It is interesting to note that the dS Swampland conjectures also lead to the solution of the `Why Now' problem in cosmology, i.e., why do we live just at the time when dark energy has become dominant, by relating the dS lifetime, which sets the scale for a typical time scale in dS, to $\Lambda^{-1/2}$ (up to possible logarithmic corrections).}, where the matter/radiation equality temperature coincides with the temperature where the dark energy begins to dominate.   In particular we do not need to appeal to Weinberg's anthropic argument \cite{Weinberg:1987dv}, which offers another explanation of this coincidence.  Weinberg's argument is that the matter dominated era should happen before dark energy takes over so that galaxies can form.  The value of the cosmological constant is chosen by avoiding any fine tuning on its value, other than the requirement of structure formation, which requires $\Lambda^{1/4}\leq T_{MR}$.  A flat prior leads to $\Lambda^{1/4}\sim T_{MR}$.   In other words, Weinberg's argument directly connects the anthropic principle to an explanation of this cosmological coincidence. By contrast in our setup the cosmological coincidence is automatic and does not appeal to anthropics. 

The idea that some components of the dark matter decay or get lighter at the Hubble scale, as is the case in our scenario, may automatically lead to a reduction of the $H_0$ tension, as proposed in \cite{Pandey:2019plg,Vattis:2019efj,Agrawal:2019dlm}.  This is an interesting link worth further study.

The organization of this paper is as follows:  In section 2 we spell out the basic idea of this model, and derive general features of the dark dimension cosmology.  More detailed derivation is discussed in section 3.  In section 4 we conclude the paper and present some topics for future research.

\section{Basic Idea}

Let us start with the basic setup:  A quasi-dS space of dimension $d$ with a small cosmological constant $\Lambda$ and $n$ mesoscopic dimensions of length scale $\Lambda^{-1/ d}$.  As discussed in \cite{Montero:2022prj}, our universe corresponds to the case of $d=4$, $n=1$ and $\Lambda=10^{-122}$ in Planck units, but for the discussions in this section it is instructive to keep $d$ and $n$ general.
We would like to study the cosmology in this setup, where the mesoscopic and microscopic dimensions are stabilized and only the d-dimensional space-time is undergoing expansion.
Moreover in this scenario the light fields of the Standard Model, except for the graviton are localized on a brane in the mesoscopic dimension.

Let $M_{\mathrm{min}}$ denote the minimum of the masses of the moduli fields fixing the internal geometry including that of the SM brane.
To keep the extra dimensions from interfering with the $d$-dimensional cosmology we must require that 

\begin{equation}
    T<M_{\mathrm{min}}\label{sh1}
\end{equation}
We thus start the universe at a temperature for the SM fields satisfying \eqref{sh1}. This, however, is not enough to be compatible with observations: we will assume that only the SM brane fields are in thermal equilibrium at this temperature, and that the bulk modes of the mesoscopic dimensions are essentially unpopulated initially. 
If we have thermal equilibrium between bulk fields and the fields on the brane near the BBN era, this would lead to too many light fields (since the light KK modes of the bulk would be excited and as we will see they would not have decayed by the Hubble time) which would be in contradiction in the scenario where our universe has $N_\mathrm{eff}\sim 3$ at the $T\sim$ MeV.  So we assume the initial conditions are such that only the brane modes, i.e. the Standard Model fields are excited and in equilibrium at temperature $T_i>$ MeV, after which it starts cooling off. 

Once this setup is fixed, so is its dynamics as we will see.   There are two basic steps: the production of dark gravitons, the KK modes in the bulk, as in \cite{Arkani-Hamed:1998sfv}.  And the second step which is the decay of the dark gravitons, which predominantly reshuffles among the dark graviton states\footnote{These decays can include decays to other KK towers in the bulk as well, which does not change our conclusions as the amplitudes are basically the same.} without appreciably changing the total mass in the dark sector.
This type of situation has been considered in \cite{Dienes:2011ja}.  Thus to study the total energy density in the dark sector, and if we are not interested in the details of the composition of the dark sector as it evolves in time, we only need to focus on the first step, i.e., the production of the dark gravitons, as we will do in this section.  Here we will perform a rough estimate of the dark matter energy density and in the next section we study this in more detail as well as the composition of dark matter and how it evolves over time.

Due to the 5d equivalence principle, the bulk graviton $h_{\mu\nu}$ couples to the localized brane modes universally as
\begin{equation}
    \label{eq:SMinteraction}
  \frac{1}{{\hat M}_p^{(n+d-2)/2}}\int d^dx\, h_{\mu\nu}(x,z=0)\,  T^{\mu\nu}(x),
\end{equation}
where ${\hat M_p}$ is the $n+d$ dimensional Planck scale and $z$ denotes the mesoscopic dimension, $z=0$ being the position of the Standard Model brane and $T^{\mu\nu}$ the energy momentum tensor for the fields on the brane.  This interaction leads to the production of dark gravitons, which are the KK excitations of the graviton in the mesoscopic dimensions, as the brane cools off.  Using dimensional analysis we learn (ignoring the expansion of the universe) that the rate at which the dark graviton energy density is produced when the brane modes are at temperature $T$ is given by (see also \cite{Arkani-Hamed:1998sfv})
\begin{equation}
  \left.\frac{d\rho_{KK}}{dt}\right\vert_{\text{production}}\sim \frac{T^{n+2d-1}}{\hat M_p^{n+d-2}}.  
\end{equation}
We will assume that the universe starts in a radiation-dominated era, as our own. During this epoch, the energy density redshifts as $a(t)^{-d}$ in terms of the scale factor $a(t)$, simply due to the expansion of the universe, and so does the temperature. The bulk KK modes that we produce essentially redshift as matter, since they are produced thermally and thus the distribution of KK gravitons follow some sort of graybody spectrum, so the masses that are populated are those with $m\sim T$. 
Because the KK modes redshift as matter, it is convenient to define $y={\rho_{KK}/s}$, where $s\sim T^{d-1}$ scales as the entropy in the light modes. This quantity removes the cosmological expansion effects and would be conserved in absence of production/decay. We start the cosmology in $d$ dimensions in the radiation dominated era where $T\sim M_p  (M_p t)^{-2/d}$, and learn that the total $y$
generated until the final $T_f$ is insensitive to $T_f$ and is dominated by the initial temperature:
\begin{equation}
  y\sim \frac{T_i^{n+\frac{d}{2}}}{{\hat M_p}^{n+d-2}M_p^{1-\frac{d}{2}}}  
\end{equation}
This result is valid only if the total mass of the dark KK gravitons remains approximately constant over time. We will be assuming this is the case, and we confirm this is the case in the next section.

As explained in \cite{Montero:2022prj}, in the Dark Dimension scenario, the cosmological constant is related to the mass scale of the tower of states (the KK scale) as
\begin{equation} \Lambda\sim m_{KK}^d\quad \Rightarrow\quad V_n=m_{KK}^{-n}=\Lambda^{-\frac{n}{d}}.\end{equation}
Upon using $M_p^{d-2}={\hat M_p}^{n+d-2}V_n={\hat M_p}^{n+d-2}\Lambda^{-{\frac{n}{d}}}$ we learn that
\begin{equation}
\label{yeq}
    y\sim \frac{T_i^{{n+\frac{d}{2}}}}{ \Lambda^{\frac{n}{d}}{M_p^\frac{d-2}{2}}},
\end{equation}
 and $y$ encodes the energy density of dark matter; we are particularly interested in the temperature $T_{MR}$ where the contribution of dark matter and that of the thermal bath living in the Standard Model brane become equal\footnote{We are assuming here that most of the matter is made of dark matter, which is a good approximation.}, the so-called matter-radiation equality temperature:
\begin{equation}
    T_{MR}\sim y,
\end{equation}
which heavily depends on the initial temperature $T_i$. The temperature of matter-radiation equality is well determined from cosmological observations \cite{2013ApJS..208...19H,Planck:2018vyg} and we could use its value to find the corresponding initial temperature $T_i$ using \eqref{yeq}:
\begin{equation}
\label{yeq2}
    T_{MR}\sim \frac{T_i^{{n+\frac{d}{2}}}}{ \Lambda^{\frac{n}{d}}{M_p^\frac{d-2}{2}}}.
\end{equation}
This is the standard way one could proceed in phenomenology.
However, we will now see that we can do better, and motivate the correct range for $T_i$ using Swampland principles.

We do not wish to make any detailed assumptions about how the universe got to the particular initial state where only the SM brane is thermal at temperature $T_i$. However, it is natural to expect higher energy density in prior epochs, and in particular, that the bulk moduli fields such as the radion and other moduli controlling the ``width'' and shape of the SM brane in the mesoscopic and microscopic dimensions might have been excited. Some of these fields decay to fields in the Standard Model brane via couplings
\begin{equation}
   \frac{1}{ {M_p}^{(d-2)/2}}\int \phi\, {\cal L}_d  \label{lag0}
\end{equation}
Where $M_p$ denotes the d-dimensional Planck mass.  In addition 
there are moduli fields that couple to the dark sector but not directly to the SM brane.   Since all these scalars affects the size of the extra dimension and geometry of SM brane, whenever they are not stabilized one may expect that the physics is effectively not four-dimensional. 

These scalars will decay by dumping their energy to the standard model fields and gravitons. The coupling to a single mode will be given by \eqref{lag0}, so that a lower bound on the decay rate of these scalars is\footnote{Note that unlike massive KK modes, the moduli fields, have effectively no KK momentum and approximate conservation of KK momentum in the mesoscopic dimension will not lead to additional KK decay channels.}
\begin{equation}
\label{decayd}
\Gamma_{\text{decay}}\sim \frac{m_\phi^{d-1}}{M_p^{d-2}}
\end{equation}
To have a meaningful four-dimensional phase, one should wait several inverse decay widths $\Gamma_{\text{decay}}^{-1}$.
On the other hand the lifetime of dS is expected to be of order of the Hubble scale (up to logarithimic corrections), based on both the dS Swampland conjecture~\cite{Obied:2018sgi,Garg:2018reu,Ooguri:2018wrx} as well as the TCC~\cite{Bedroya:2019snp}.  Taking both of these things together, the existence of a quasi-dS phase means that the decay rate of the moduli fields should be bigger than the Hubble scale,
\begin{equation}
  \Gamma_{\text{decay}}>M_p \cdot \left(\frac{\Lambda}{ M_p^d}\right)^{1/2}  
\end{equation}
which leads to

\begin{equation}
  m_\phi > M_p \left(\frac{\Lambda}{ M_p^d}\right)^{\frac{1}{2(d-1)}}  
\end{equation}

This implies that a safe initial temperature to have where the moduli field would have decayed and settled to their minimum and that would not be further produced is 

\begin{equation}
  T_i\sim  M_p \left(\frac{\Lambda}{ M_p^d}\right)^{\frac{1}{2(d-1)}} <m_\phi  \label{eqc1}
\end{equation}

Plugging this value for initial temperature we find
\begin{equation}
    y\sim {T_i^{{n+{\frac{d}{2}}}}\over  \Lambda^{{n\over d}}{M_p^{d-2\over 2}}} \sim M_p\left({\Lambda \over M_p^d}\right)^{{d^2-2nd+4n \over 4d(d-1)}}
\end{equation}

Note that, as already mentioned $y\sim T_{MR}$, i.e., the temperature at which the dark graviton density is equal to radiation density.  If we assume that the bulk of the matter is in dark graviton modes, requiring that this temperature is close to the temperature at which dark energy dominates we would require
\begin{equation}
T_{MR}\sim \Lambda^{1\over d}
\end{equation}
This leads to
\begin{equation}
  d^2-2nd+4n=4(d-1) \rightarrow  n={\frac{d}{2}}-1
\end{equation}
In other words for $n$ satisfying this relation, the matter radiation density equality temperature is near the temperature where the dark energy dominates.
This includes our universe in the dark dimension setup where $d=4,n=1$!
We have thus explained the coincidence
problem in the dark dimension scenario, with dark gravitons playing the role of dark matter\footnote{Note that for the matter/radiation equality to happen before the dominance of dark energy we must have $n\ge \frac{d}{2}-1$; on the other hand for massive gravitons to be the bulk of the matter in our scenario requires $T_{MR}<T_{i}$ which leads to ${\frac{d}{2}}> n$.  Thus in our scenario, assuming $n$ is an integer, the only option for matter/radiation equality to happen before dark energy dominance is for the coincidence to also happens, i.e. for $n={\frac{d}{2}}-1$}, and assuming that all of the inequalities above are saturated or at least set the scale for $T_i$. Notice that saturation of the inequalities also predicts the existence of a massive modulus at a scale near $\sim\Lambda^{1/6}\approx 1$ GeV.

It is amusing to note that if we require $d+n\leq 11$ (to embed in M-theory) and $n$ to be integer\footnote{Note that for odd $d$ we can interpret the fractional values of $n$ as an internal space with anisotropic extra dimensions.}, the only options for the coincidence to take place are $(d,n)=\{(8,3),(6,2),(4,1),(2,0)\}$. More generally we have seen that if we start with the temperature (in Planck units) of $T_i \sim \Lambda^{\frac{1}{2(d-1)}}$ the cosmological coincidence of the matter/radiation/dark energy densities will automatically happen at $T=\Lambda^{1\over d}$ if $n={\frac{d}{2}}-1$ with the bulk of the matter in dark graviton with mass initially near $T_i$. 

In the next section we expand upon this rough derivation and fill in some details for the specific case of $d=4,n=1$ in our universe, where we will find $T_i\sim 4$ GeV.

\section{Dark Dimension Cosmology: Production and abundance of KK gravitons}
In this section we restrict our attention to the dark dimension scenario relevant for our universe where
$d=4,n=1$. 

\subsection{General derivation}

In the previous section we showed how the dark dimension scenario naturally produces the right abundance of dark matter $\rho_{\text{DM}}$, simply by starting the brane at an initial temperature $T_i$ (which saturates interesting Swampland inequalities) and letting it evaporate into the bulk. But a successful dark matter model must get a few other things also right.  In addition to the correct dark matter density $\rho_{\text{DM}}$, the lifetime of the dark matter to decay to standard model fields and in particular photons must be much larger than the current age of the Universe.  Moreover there are strong bounds on how much dark matter energy density can change over time.
Therefore, we must analyze the decays of KK gravitons. One possible decay channel, forced on us by the production mechanism, is back to the SM brane. The decay rate of a massive KK graviton of mass $m$ to a massless (light) field is given by dimensional analysis
\begin{equation}
\label{dark}
  \Gamma_{KK}\sim \lambda^2 {m^3\over M_p^2}
\end{equation}
This sets the decay rates to the SM brane. This is comparable to the age of the Universe for $m\sim 100$ MeV, so setting the temperature below this scale would seem to be fine. However, \cite{Slatyer:2016qyl} shows that most of the dark matter must have a lifetime of order $\sim10^7$ times the age of the Universe, or otherwise its decays to photons would generate a much too large CMB anisotropy. This translates to $m\lesssim 1$ MeV, dangerously close to BBN. But if we set the initial temperature to be this low, we do not get the right dark matter abundance, as shown in the calculation of the previous section.

Luckily, there are many additional decay channels available for the KK gravitons: intra-tower decays, which is indeed a more dominant decay mode. In the absence of isometries in the mesoscopic dimension, which is the generic expectation, the KK momentum of the dark tower is not conserved. There will be decays where one dark graviton of KK quantum $n$ decays to two other ones, with quantum numbers $n_1$ and  $n_2$.  The decay rates for each channel is expected again to be roughly given by \eqref{dark}.  If the KK quantum violation can go up to $\delta n$, the number of channels available is $\sim {m\over m_{KK}}\times \delta n$.  Taking into account the phase space factor (which, since the decay is almost at threshold, is roughly the velocity of decay products, $\sim( m_{KK} \delta n /m)^{1/2}$ leads to 
\begin{equation}
\label{totd}
    \Gamma_d^{\text{tot}}\sim \beta^2 {m^3\over M_p^2} {m\over m_{KK}}\times \delta n \times\sqrt{m_{KK}\delta n \over m}\sim \beta^2(\delta n)^{3/2}\frac{m^{7/2}}{M_P^2m_{KK}^{1/2}}
\end{equation}
where $\delta n$ is the maximal range of KK violation and $\beta$ parametrizes our ignorance of decays in the dark dimension (see also below Eq.~\ref{e2334} for more details).  This rate typically overwhelms \eqref{dark}.   The typical mass loss in the decay is
$\delta m\propto m_{KK}\delta n$.
We will find that to be consistent with observation $\delta n$ cannot be large and is in the range $1\leq \delta n\leq 10^2$. Note that $\delta n$ is equivalent to saying that that the dark dimension is smooth and homogeneous at scales smaller than $(m_{KK}\delta n)^{-1}$, and from this perspective it is reasonable that $ \delta n$ is not too large.\footnote{ The more refined version of \eqref{totd} can be obtained along the lines of \cite{Mohapatra:2003ah}:  the violation in KK mode number can be viewed as having the $n$-th mode of a dimensionless field $g_n$ in the dark dimension having a vev of order one.  Expanding the Einstein-Hilbert action with these vevs leads to a 4d term (suppressing the tensorial structure)
\begin{equation} {1\over M_p}\int d^4x\ g_n\ h^*_k(\partial h_l) (\partial h_{k-l-n})
\end{equation}

The decay rate for a mass $m$ KK mode to two of masses $m_1$ and $m_2=m-m_1- n\, m_{KK}$  (assuming $m\gg n\ m_{KK}$) yields
\begin{equation} 
\Gamma_d\propto \frac{\beta^2}{M_P^2} \left(\frac{\sqrt{m_1(m-m_1)}(m^2 + m_1^2 - m m_1)^2}{m^{5/2}}\right) \sqrt{m_{KK} n}.
\label{drt}
\end{equation}
 
The total decay rate of the mode of mass $m$ is obtained by summing the above over all possible decay channels, and goes as \eqref{totd}.}

To sum up, even though the composition of the dark matter is changing and it is getting reshuffled to lighter mass states, the total mass in the dark sector does not change appreciably from when they were produced as we will shortly show. 
What we are finding here is a specific realization of the dynamical dark matter scenario proposed in \cite{Dienes:2011ja}.

We can also estimate the typical dark matter mass as a function of time, simply by noticing that at times larger than $1/\Gamma_d^{\text{tot}}$, where $\Gamma_d^{\text{tot}}$ is given by \eqref{totd}, dark matter heavier than the corresponding $m$ has decayed. Therefore we expect the typical dark matter mass to be at the edge of range which is about to decay
(which would be valid for $t\gtrsim t_i$, with $t_i=1/\Gamma^{\text{tot}}_i$, and $\Gamma^{\text{tot}}_i$ the total decay rate at $m\sim T_i$)
\begin{equation} 
\label{matt}
m_{\text{DM}} \sim \left(\frac{M_P^4 m_{KK}}{(\delta n)^3\beta^4}\right)^{\frac17}\frac{1}{t^{\frac27}}.
\end{equation}

\begin{figure}
	\centering{}
	\includegraphics[scale=0.4]{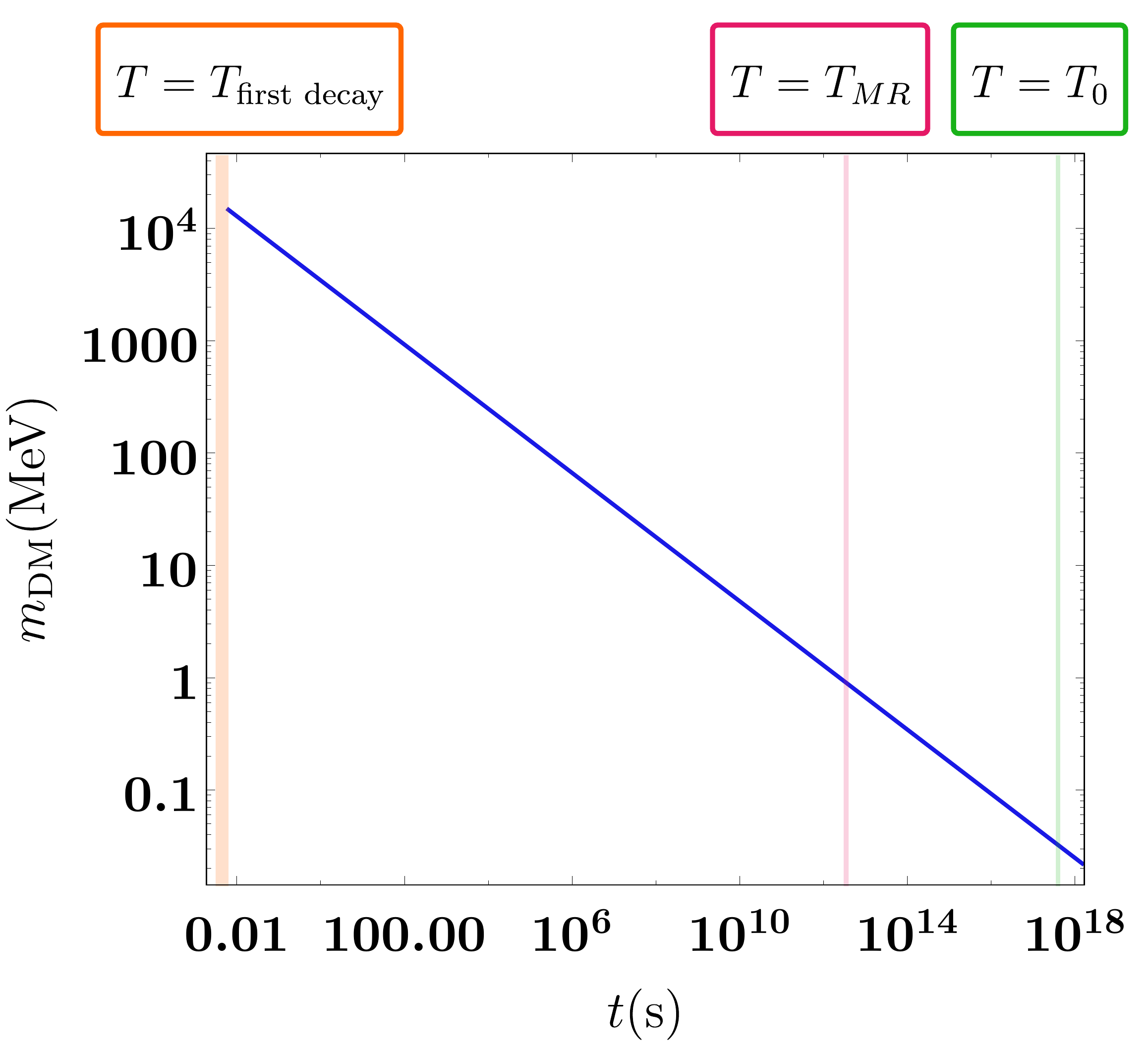}
	\caption{\footnotesize Analytic formula for the dark matter as a function of time for the choice $\delta n=100$ and $\beta=20$. The orange vertical shows the lifetime of the initial dark graviton distribution to decay into lighter states of the dark tower. $T_{MR}$ denotes the matter/radiation equality temperature and $T_0$ is the temperature today.  In Section 3.2 we verify this behaviour by a more detailed analysis.}
	\label{MRanalytic}
\end{figure}

See Figure \ref{MRanalytic} for how DM mass changes with time using this formula, for a sample value of $\delta n \sim 10^2$ and $\beta \sim 20$.
We can estimate the fractional rate of change in total mass of the dark sector using the fact that each decay loses  $\sim \delta n \ m_{KK}$ in mass (ignoring the usual $-3H$ cosmological dilution):
\begin{equation}
\label{dmdecay}
\frac{1}{ \rho_{\text{DM}}}\frac{d\rho_{\text{DM}}}{dt}\sim -\Gamma^{\text{tot}}_d(m_\mathrm{DM}) \frac{\delta n \ m_{KK}}{ m_{\text{DM}}}\sim -H \frac{\delta n\ m_{KK}}{m_{\text{DM}}}, \end{equation}
where we used the fact that $\Gamma^{\text{tot}}_d$ and $H$ are both proportional to $1/t$.
This is typically small because in all epochs ${\delta n\ m_{KK}\over m_{\text{DM}}}\ll 1$ (the biggest value is achieved now, when $m_{\text{DM}}$ is the smallest).  This confirms that the loss of mass in the dark sector is rather small and all that happens is that the composition of the dark matter sector become lighter as the time evolves.

We can also estimate the profile of the dark matter mass distribution.  The number density of the dark matter spans from $m_{KK}\sim 1$ eV to the upper range $m_{\text{DM}}$
\begin{equation} 1 \ \text{eV}< m< m_{\text{DM}} \end{equation}
and its mass density is peaked near $m_{\text{DM}}$.  Moreover the mass density drops off after $m_{\text{DM}}$ due to exponential suppression which goes as $\sim \exp(-(m/m_{\text{DM}})^{7/2})$.

To be consistent with \cite{Slatyer:2016qyl}, we need that the lifetime of the decay of matter to the photons is bigger than $10^7-10^8$ times the life of the universe between the time of CMB formation until the reionization.  To get such a long lifetime, the mass of the dark matter during that epoch should be bounded above by 10-1000 keV during this period.  On the other hand the mass of the dark matter from the time of matter/radiation equality till today decreases \eqref{matt} by a factor of $(t_{\text{today}}/t_{MR})^{2/7}\sim 25$, which means that 
\begin{equation}m_{\text{DM}}(\text{today})\sim 1-100\  \text{keV}\end{equation}
In principle this could have been an upper bound of the DM mass today.  However,
the mass of the dark matter today cannot be much less that 1 keV because otherwise we would be getting too rapid a decay of the dark matter density, using \eqref{dmdecay}.  So observations limit us to a narrow allowed range.
The question is whether this rather narrow range for the allowed DM mass  comes out naturally from our model.  Indeed if we use \eqref{matt} we see that for $\beta \sim 20 $, $\delta n\sim 10^2$ we get

\begin{equation}m_{\text{DM}}(T_{\text{MR}})\sim 1 \ {\rm MeV},
\label{1Mev}
\end{equation}
and
\begin{equation}m_{\text{DM}}(\text{today})\sim 50  \ {\rm keV}.
\label{1kev}
\end{equation}

Thus, we have a well motivated dark matter candidate in the dark dimension scenario which passes all the experimental checks.

\subsection{A numerical approach}
We will now show that the estimates above are quite accurate, by solving in detail for the abundances of the different particles in the tower, taking into account production from the SM brane, their decay to degrees of freedom on the SM brane, as well as the intra-KK tower decays. This is achieved by means of a Boltzmann equation (see e.g.~\cite{Dodelson:2003ft}). We begin by considering a tower of equally spaced dark gravitons, indexed by an integer $l$, and with mass $m_l=l\, m_{KK}$. We denote by $n_l$ the physical number density of species $l$. There is a natural dilution of $n_l$ due to the fact that the universe expands. Since
\begin{equation} \dot{n_l}+3 H n_l = \frac{1}{a^3}\frac{d(a^3 n_l)}{dt} = T^3 \frac{d}{dt}\left(\frac{n_l}{T^3}\right)
,\end{equation}
where in the last step we neglected the time derivative of $g_{*}(T)$, the number of relativistic\footnote{At this level of precision, we also neglect the difference between the entropy degrees of freedom and the relativistic degrees of freedom.} degrees of freedom. To remove this expansion term, we write everything in terms of $Y_l= \frac{n_l}{T^{3}} $, which does not dilute. The equation of evolution of $Y_l$ is then
\begin{equation}\dot{Y}_l=\Gamma_{\text{SM},l} -  (\Gamma_{l,\text{SM}}+ \Gamma^{\text{tot}}_l) Y_l +2  \sum_{l'>l} \Gamma_{l'\rightarrow (l,l'-l)} \ Y_{l'},\label{beq}\end{equation}
where the different decay rates in the above equation are as follows:\begin{itemize}
    \item $\Gamma_{\text{SM},l}$ is the production rate of dark gravitons per unit volume in mode $l$ via collisions on the SM brane. These can be computed from the universal coupling of matter fields to the bulk gravitons \eqref{eq:SMinteraction} upon expanding in the KK graviton modes
    $$h_{\mu\nu}(x,z)=\sum_l h_{\mu\nu}^l(x) \phi_l(z)$$
    (where the $l=0$ term is the massless graviton),
    and leading to 
  \begin{equation}
      \sim \frac{1}{M_p}\sum_l\int d^4x\, h_{\mu\nu}^l(x)  T^{\mu \nu}(x).
  \end{equation}
    The contribution to the production rate of KK gravitons from massless particles at temperature $T$ was computed using this interaction in \cite{Hall:1999mk}:
    \begin{equation} \Gamma_{\text{SM},l}=\frac{\lambda^2 T m_{KK}^5}{128\pi^3 M_P^2}K_1(l m_{KK}/T)\, l^5,\end{equation}
    where $K_1$ is a Bessel function of the first kind, $T$ is the temperature of the SM brane, and we introduced the factor $\lambda$ in order to account for the number of degrees of freedom of the SM brane, as well as for the overlap between the bulk graviton wavefunctions and the SM brane.  We need to include the appropriate (spin dependant) contribution of every SM particle with a mass below the initial temperature ($T\sim 4$ GeV). In principle, $\lambda^2$ depends on time but the production is only relevant at temperatures close to the initial temperature. For the plots, we consider as massless all particles up to the tau lepton which gives an estimate $\lambda^2 \sim 56$.  
    
    \item $\Gamma_{l,\text{SM}}$ is the decay rate of bulk gravitons back to the SM brane. Again, following \cite{Hall:1999mk}, we have\begin{equation} \Gamma_{l,\text{SM}}=\frac{\tilde{\lambda}^2m_{KK}^3}{80\pi M_P^2}l^3.
    \label{decaytophoton}
    \end{equation}
    where $\tilde{\lambda}$ similarly takes into account all the available decay channels and is a function of time. However, we find that this decay channel is irrelevant for determining the dark matter distribution and its total abundance and any $\mathcal{O}(1)$ choice of $\tilde{\lambda}$ does not change our results. On the hand, dark matter decays to photons provide strong constraints~\cite{Slatyer:2012yq, Slatyer:2016qyl} (see also \cite{Ibarra:2013cra} for a review of bounds on decaying DM) and in that case the appropriate choice is $\tilde{\lambda}^2 \sim 1$.
    \item $\Gamma_{l'\rightarrow (l,l'-l)}$ is the fundamental quantity controlling the decays within the dark tower, which we computed in \eqref{drt}. It is the decay width for the particle at level $l'$ to decay to particles at levels $l$ and $l'-l$ (where we have assumed that the violation of KK momentum is small). In terms of $l$, it reads, after summing over the possible KK momentum decay channels, 
    \begin{equation}
    \Gamma_{l'\rightarrow l}
    =\frac{1024}{85\pi}\beta^2 \frac{m_{KK}^3}{M_P^2}\frac{\sqrt{l(l'-l)} (l^2 - l l' +(l')^2)^2}{(l')^{5/2}} (\delta n)^{3/2}
    \label{e2334}\end{equation}
    We have introduced a factor of $\beta$ to parametrize our ignorance about the dark dimension decays. It absorbs the expectation value of the vevs of the KK modes parametrizing inhomogeneities in the dark dimension, as well as the triple overlaps between KK modes and the number of towers available to decay.  Here we have called the dark graviton the dark matter, but it should be understood that there could easily be other light modes in the bulk which lead to new decay channels and dark matter components.  In particular we do expect dark fermions to propagate in the bulk playing the role of sterile neutrinos, as discussed in \cite{Montero:2022prj}.  The existence of these other components of KK modes does not affect the total abundance of DM mass as they just distribute among each other, as they decay from one to another.  We expect all decay KK channels to be captured by the above formula. In particular when we say dark graviton is the dark matter, we mean also possibly including these additional KK towers.
    The existence of these channels leads us to expect that perhaps $\beta >1$.
    
Finally, there is a factor of 2 in the term multiplying $\Gamma_{l'\rightarrow l,l'-l}$ in \eqref{beq}. This is due to the fact that a particle at level $l$ can appear as either of the two decay products at the particle at level $l'$; the contribution is therefore doubled.
    \item Finally, $\Gamma^{\text{tot}}_l$ is the sum of \eqref{e2334} over all decay channels:
        \begin{align}
        \Gamma^{\text{tot}}_l
        &=\sum_{l'}\Gamma_{l\rightarrow l'}
        =\frac{1024}{85\pi}\beta^2 \frac{m_{KK}^3}{M_P^2} (\delta n)^{3/2} \sum_{l'<l} \frac{\sqrt{l'(l-l')} (l^2 - l l' +(l')^2)^2}{(l)^{5/2}} \nonumber \\
        &= \beta^2 \frac{m_{KK}^3}{M_P^2} (\delta n)^{3/2} \, l^{\frac72}.
        \label{e23354}\end{align}
\end{itemize}

\begin{figure}
	\centering{}
	\includegraphics[scale=0.5]{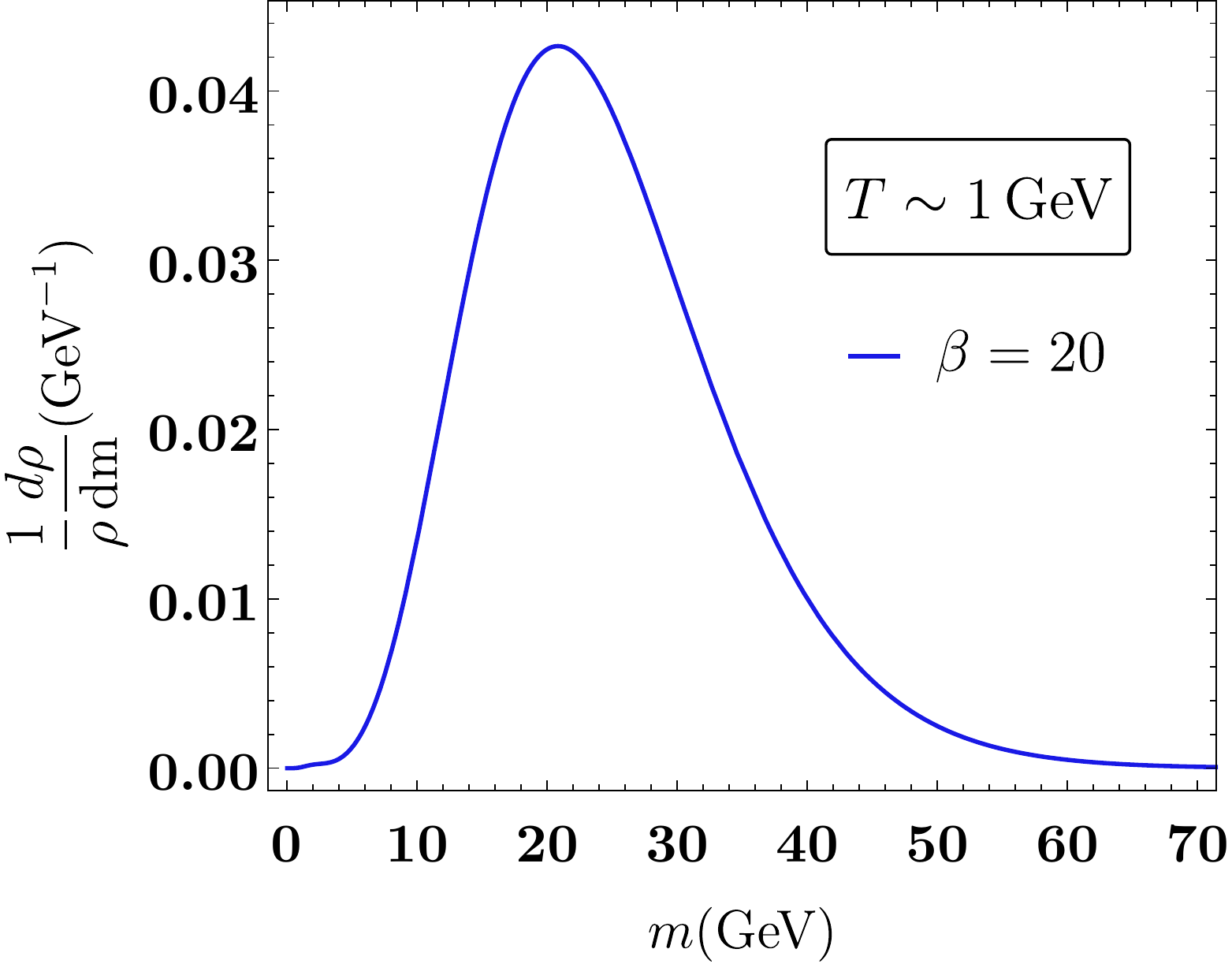}
	\caption{\footnotesize Distribution of the energy density of dark matter as a function of the mass of the KK modes at a temperature $T\sim 1$ GeV far enough below the initial temperature $T_i\sim 4$ GeV so that production by photons has had time to take place, but close enough so that decays are negligible. We normalize by the total dark matter energy density, so the integral under the curve is one.}
	\label{initial}
\end{figure}

\begin{figure}
	\centering{}
	\includegraphics[scale=0.5]{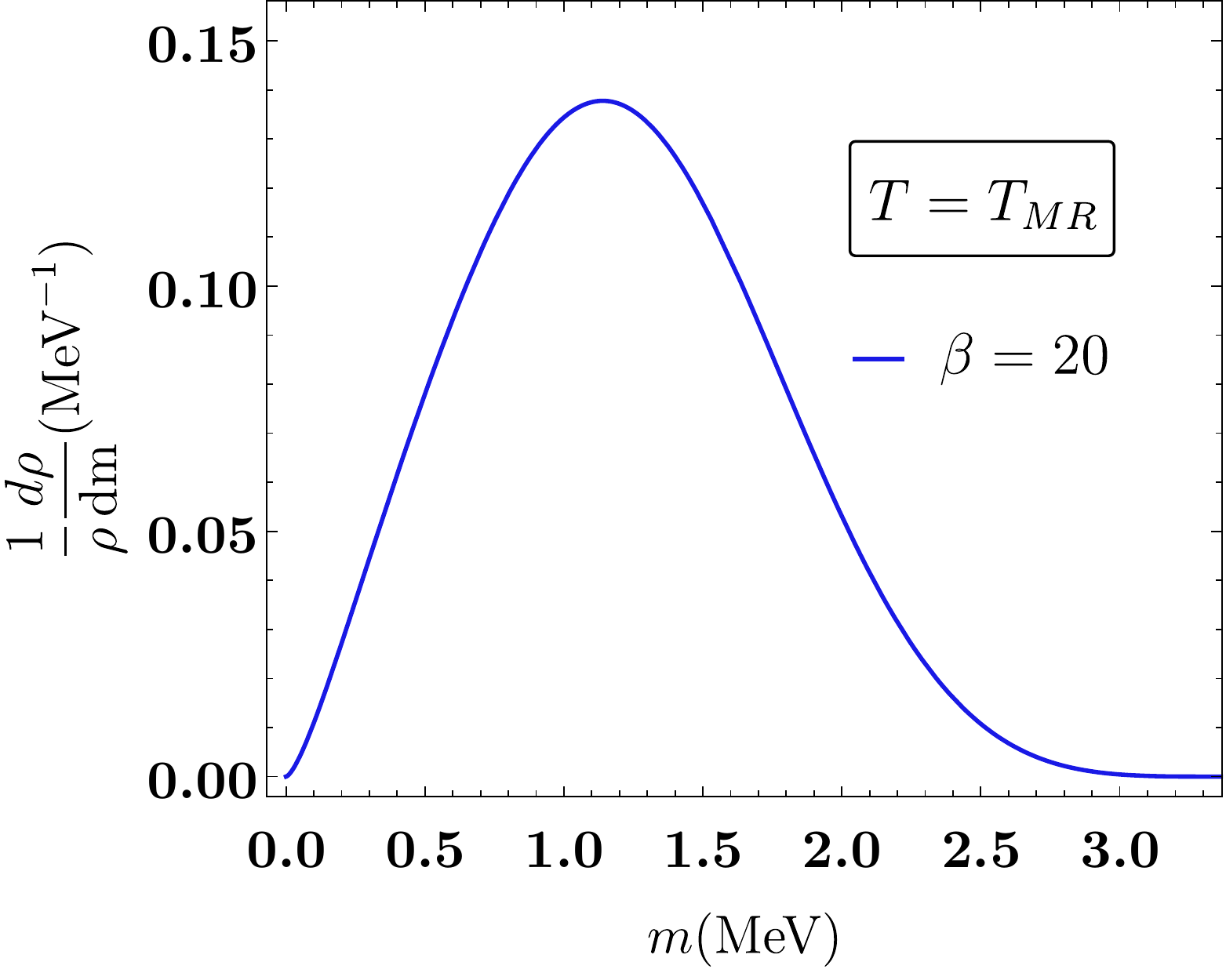}
	\caption{\footnotesize Distribution of the energy density of dark matter as a function of the mass of the KK modes at a temperature $T \sim 1$ eV, which is the matter/radiation equality temperature. We divide by the radiation energy density, and since we evaluate at equality, the integral under the curve is one.}
	\label{MR}
\end{figure}

\begin{figure}
	\centering{}
	\includegraphics[scale=0.5]{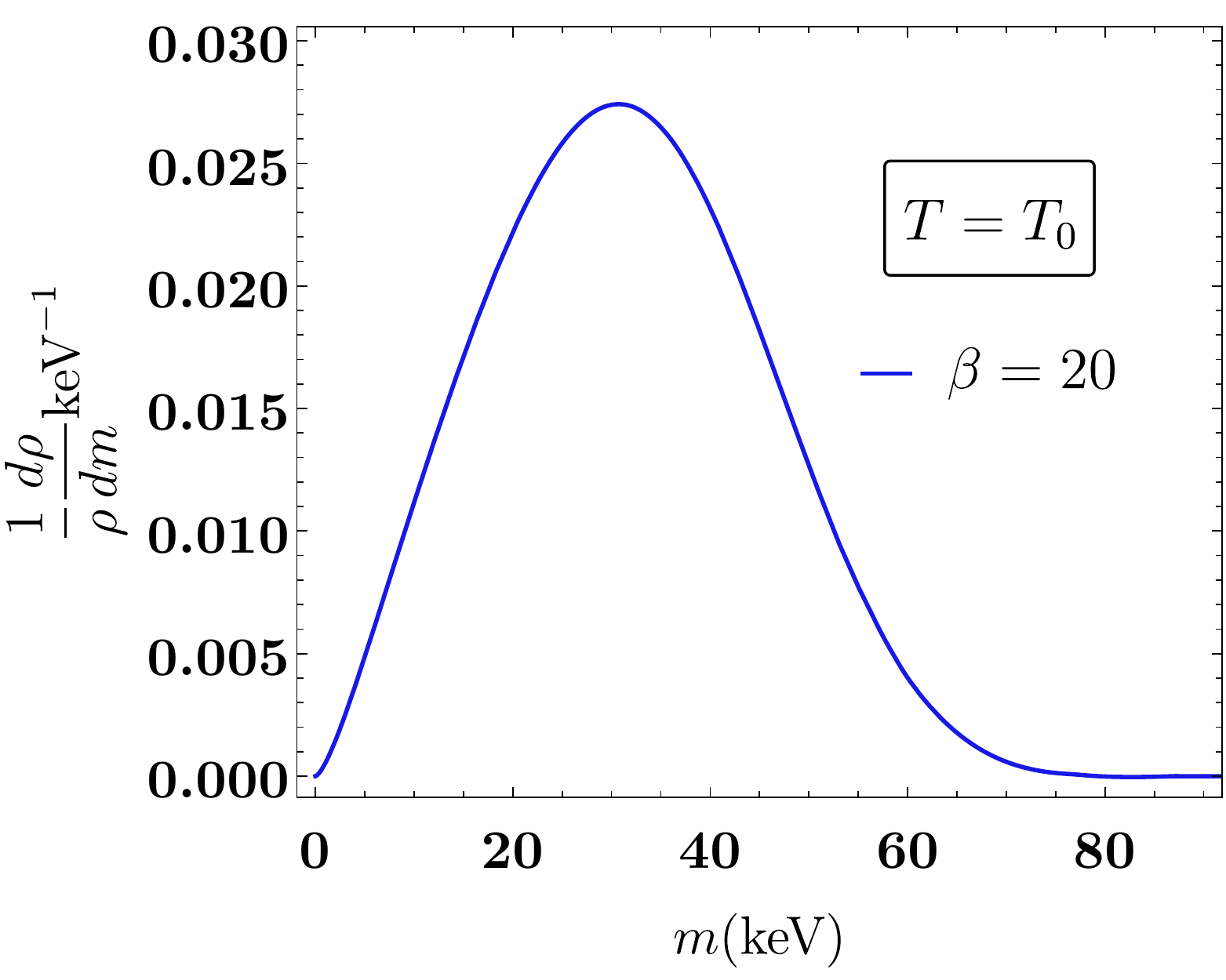}
	\caption{\footnotesize Distribution of the energy density of dark matter as a function of the mass of the KK modes at a temperature $T \sim 0.2$ meV, which is the temperature today. We divide by the total dark matter energy density so the integral under the curve is one.}
	\label{present}
\end{figure}

In Figures \ref{initial}, \ref{MR} and \ref{present} we present the results of numerically solving \eqref{beq}, for the choice of parameters $\beta=20$, $\delta n = 100$, $T_i = 4$ GeV. In the figures we find the mass distribution of dark matter at different times. The initial distribution produced by collisions on the brane is shown in Fig. \ref{initial}. The position of the peak is above the initial temperature by a factor of $\sim 5$ which is the maximum of the product of a polynomial and the modified Bessel function $K_1$ as functions of $\frac{m}{T_i}$. The reason for this is that modes a bit more massive than $T_i$, even though they are not fully produced, are the dominant ones in the energy density, since their mass is higher. At times very close to the initial temperature we can neglect all of the decay terms in the Boltzmann equation. Changing variables from time to temperature and neglecting the term coming from derivatives of the number of relativistic degrees of freedom we can solve the equation to find that the contribution of a mode of mass $m_l$ is given by
\begin{equation}
    \left.\frac{d\rho_{\text{dm}}}{\rho_r}\right\lvert_{T_f} = \lambda^2 \frac{315 \sqrt{10}m^2 dm}{64 \pi^{6} T_f M_p m_{KK}} \int_{\frac{m}{T_i}}^{\frac{m}{T_f}} \frac{u^3 K_{1}(u) du}{g_{\star}^{3/2}(u)}.
    \label{distributioneq}
\end{equation}

We have checked that this curve nicely matches our numerical solution. Integrating over the mass we find an expression for the dark matter energy density over the energy density in the radiation:
\begin{equation}
    \left.\frac{\rho_{\text{DM}}}{\rho_r}\right\lvert_{T_f} = \lambda^2 \frac{1890 \sqrt{10}}{\pi^{6}T_f M_p m_{KK}} \int_{T_i}^{T_f} \frac{T^2 dT}{g_{\star}^{3/2}(T)} 
\end{equation}

Taking into account the contribution from baryonic matter and using the data from \cite{Planck:2018vyg}, this ratio should be equal to $\sim 0.6$ for $T_f\sim 0.8$ eV. This fixes the initial temperature to about $4$ GeV, which matches the numerical result.

In Fig. \ref{MR} we plot the distribution of DM graviton energy density at a temperature of 1 eV divided by the radiation energy density. We chose the particular value of the initial temperature so that the integral gives 1. We find that most of the dark matter particles have a mass around 1 MeV, in agreement with our previous formula \eqref{1Mev}. Using  \eqref{decaytophoton} we find that the lifetime to photons is $\tau_{\gamma}\sim 10^{24}$ s. This illustrates the point that dark gravitons satisfy stability bounds on the lifetime of dark matter. They are constantly decaying (but their decays essentially conserve mass) and by the time we reach matter domination their lifetime to decay back into the brane is large enough to be consistent with current bounds \cite{Slatyer:2016qyl}. We have checked that production of photons is negligible at all times. More concretely, we checked that the fraction of energy which is being transferred back into the brane is always negligible. We also checked that the total fraction of energy lost in the decays (which is a dominant effect with respect to energy lost by decaying back into the brane) is less than 1 per cent. We verified that after production the total dark matter energy density falls exactly as $T^3$, as it should. Finally, in Fig. \ref{present} we show the distribution at the present time. We can see that the analytic estimate \eqref{1kev} agrees well with the numerical computation. We remark that these plots correspond to a particular choice $\delta_n$ and $\beta$. As explained above, these parameters can be physically motivated only up to an order one factor. Thus, for example, the position of the peak could also change up to an order one factor. For this reason we can conclude that the mass of dark matter today lies roughly between 1-100 keV.

\section{Concluding Remarks}

    In this note we have shown how the dark dimension cosmology provides a natural, viable dark matter candidate, in the form of dark gravitons or other towers of bulk fields, produced thermally by the hot brane where the SM fields reside. The dark matter is prevented from substantially decaying back to SM fields because its mass distribution quickly shifts to lower values by the large number of available KK modes of lighter mass, leading to a small decay rate to the SM fields which scales as $m_{\text{DM}}^3$.

We have seen that there is a natural explanation of the coincidence problem in our setup without appealing to anthropics, but instead using the Swampland conditions that the mass of moduli should be heavier than Hubble scale for a quasi-dS cosmology which sets the initial temperature for cosmology.  Of course one aspect of the dark dimension scenario is that $\Lambda \ll M_p^4$.  This may require an anthropic explanation, such as requiring a large entropy for us to exist.  Indeed the entropy in our setup at the matter/radiation equality is of the order of $S\sim (\Lambda /M_p^4)^{-3/4}$ and thus large entropy requires small $\Lambda$.

The most pressing question is how to probe this scenario and distinguish it from other dark matter candidates. Of course one prediction of our setup is that a direct detection experiments of DM would not succeed, as the DM interacts only gravitationally and would be too weak to detect. Conversely, dark gravitons couple to matter much in the same way as the ordinary graviton does, and so they can be detected in principle by gravitational wave detectors. Although the characteristic frequencies are a factor of $10^8$ larger than the ones that can currently be explored by LIGO, it may be possible to detect high-frequency gravitational waves in different ways (see, e.g. \cite{Ejlli:2019bqj}).

As for specific novel aspects of our model,
one possibility that should be explored further is whether the decay of dark matter can lead to a natural explanation of the $H_0$ tension (see e.g. \cite{Agrawal:2019dlm,Vattis:2019efj,McDonough:2021pdg}), in particular our situation is rather similar to that of \cite{Agrawal:2019dlm} where the dark matter (also viewed as part of a tower) loses mass due to its coupling to a rolling scalar field.  However there are also some differences here, as we have no rolling scalars.  It would be interesting to study whether the $H_0$ tension gets reduced in our setup.

We have argued that to obtain a cosmologically successful setting with a quasi-dS space with cosmological constant $\Lambda$ the mass of the moduli of internal fields should be larger than $\Lambda^{1\over 2(d-1)}=\Lambda^{1\over 6}\sim  {\rm GeV}$ for $d=4$.  This sets the initial temperature for the cosmology
which in turn translates to the correct dark matter abundance.  So this is the analog of the ``WIMP miracle'' in our setting.  Turning this around it also suggests that perhaps the weak scale of the standard model is also set by this scale, and in particular, up to order one factors
$\langle H\rangle \sim \Lambda^{1\over 6}$.  Indeed a similar relation was noted in \cite{Montero:2022prj} where the lack of hierarchy in the neutrino sector led to this relation between the Higgs vev and the dark energy.  It would be interesting to explore the generality of this connection between the dark energy scale and the moduli field, as well as the Higgs vev. 
 This connects the issue of electroweak hierarchy to that of dark energy.
 Thus in the dark dimension scenario, we are having various mass scales all pegged to $\Lambda$, namely the Hubble scale, $\Lambda^{1\over 2}$, the KK scale
$\Lambda^{1\over d}$, the Higgs and moduli scale, $\Lambda^{1\over 2(d-1)}$ and the higher dimensional Planck scale $\Lambda^{n\over d(n+d-2)}$ (where $n$ is the number of mesoscopic dimensions). 

It is interesting to note that, if we assume that dark matter is coupled only gravitationally to our sector, as in our scenario, the experimental bounds on their decay to photons \cite{Slatyer:2016qyl} imply that their mass should be no higher than 1-100 keV.  This also gives further evidence to the lower bound on the exponent for the distance conjecture argued in \cite{Montero:2022prj} to be $\alpha > 1/4$.  If the exponent were lower it would have made the KK tower too massive to support such a light dark matter.  In particular this gives an experimental bound that the lower end of the exponent in the distance conjecture should be $\alpha > 1/(4.5)$, providing further experimental support for the bound argued in \cite{Montero:2022prj}.

Finally, we would like to note the work of \cite{Anchordoqui:2022txe} which explored the possibility of dark matter being primordial black holes in the dark dimension scenario. Although our proposal here is different, it is worth remarking that there are some relations between the two.  Namely it is found in \cite{Anchordoqui:2022txe} that the viable range for the primordial black holes to constitute all of dark matter, lead to 5-dimensional black holes whose masses are in the range $10^{14}g-10^{21}g$ corresponding to radii in the range of $R\sim (10^{-4}-1)\mu m$. In principle these PBH's can be made of KK modes with mode numbers ranging $n\sim 1-10^{4}$ leading to the KK masses in the range $1 \ {\rm eV} <m< 10 \ {\rm keV}$ rather close to the range of dark graviton tower we are finding.  Indeed this coincidence can be partially explained by the fact the decay rate of an object of mass $m$ to SM brane fields is identical (modulo phase space factors) in both approaches. This is a manifestation of the fact that, for many purposes, a black hole can be replaced by a gas of gravitons in a box \cite{Dvali:2011aa}. The two scenarios can be related if the dark gravitons that we have studied coalesce and form primordial black holes of the corresponding scale.
It is worth investigating whether this actually happens.

\vspace{0.5cm}

\textbf{Acknowledgments}\,  We thank Cora Dvorkin, Dieter Lust, Samir Mathur, Rashmish Mishra, Julian Mu\~{n}oz, Matt Reece, Luis  Ib\'a\~nez, Martin Ro{\v c}ek, Matt Strassler and Irene Valenzuela for useful discussions and comments.  In addition we would like to thank the SCGP for hospitality at the 2022 summer workshop which led to this work.

The work of MM, GO
and CV is supported by a grant from the Simons Foundation (602883, CV) and by the NSF grant PHY-2013858.  EG is supported in part by  NSF grant PHY-2013988.

\bibliographystyle{JHEP.bst}
\bibliography{ref.bib}

\end{document}